# TESLA*HERA AS LEPTON (PHOTON) – HADRON COLLIDER

**Ö. Yavaş, A. K. Çiftçi**
Univ. of Ankara, Faculty of Science, Dept. of Physics, 06100 Tandogan, Ankara, TURKEY
**S. Sultansoy**
DESY, Notke Str. 85, D-22607 Hamburg, GERMANY
Univ. of Ankara, Faculty of Science, Dept. of Physics, 06100 Tandogan, Ankara, TURKEY
Institute of Physics, Academy of Sciences, H. Cavid Ave. 33, Baku, AZERBAIJAN

*Abstract*
New facilities for particle and nuclear physics research, which will be available due to constructing the TESLA linear electron-positron collider tangentially to the HERA proton ring, are discussed.

## 1 INTRODUCTION

Construction of TESLA tangentially to HERA, will provide a number of additional opportunities to investigate lepton-hadron and photon-hadron interactions at TeV scale:

TESLA ⊗ HERA=TESLA ⊕ HERA
  ⊕ TeV scale ep collider [1-3]
  ⊕ TeV scale $\gamma$ p collider [4, 5]
  ⊕ eA collider [6]
  ⊕ $\gamma$ A collider [6]
  ⊕ FEL $\gamma$ A collider [7].

It has been shown that $L_{ep}=10^{31}$cm$^{-2}$s$^{-1}$ can be achieved within a moderate upgrade of TESLA and HERA parameters [2]. Using Compton backscattering of a laser beam off the TESLA electron beam will give the opportunity to construct a γp collider with the same luminosity and slightly less center-of-mass energy. Acceleration of nucleus beams at HERA will give opportunity to investigate e-nucleus and γ-nucleus collisions in a very promising kinematics region. An estimation shows that L·A=$10^{30}$cm$^{-2}$s$^{-1}$ can be achieved at least for light nuclei. In principle, luminosity values for all these options can be increased using a "dynamic" focusing scheme and cooling of the HERA beam both at the injectors and in the main ring. Then, colliding of the TESLA FEL beam with nucleus bunches from HERA will give an interesting possibility to investigate "traditional" nuclear phenomena. Finally, scattering of the high-energy photon beam on a polarized nuclear target can be used for an investigation of spin content of nucleons [8].

## 2 MAIN PARAMETERS OF ep COLLIDER

There are a number of reasons [1] favouring a superconducting linac (TESLA) as a source of e-beam for linac-ring colliders. First of all spacing between bunches in warm linacs, which is of the order of ns doesn't match with the bunch spacing in the HERA. Also, the pulse length is much shorter than the ring circumference. Additionally, in the case of TESLA, which use standing wave cavities, one can use both shoulders in order to double electron beam energy, whereas in the case of conventional linear colliders one can use only half of the machine, because the travelling wave structures can accelerate only in one direction. The most transparent expression for the luminosity of this collider is [9]:

$$L_{ep} = \frac{1}{4\pi} \cdot \frac{P_e}{E_e} \cdot \frac{n_p}{\varepsilon_p^N} \cdot \frac{\gamma_p}{\beta_p^*}$$

for round, transversely matched beams. Using the values of upgraded parameters of TESLA electron beam (Table I) and HERA proton beam (Table II), we obtain $L_{ep}$=1.3·$10^{31}$cm$^{-2}$s$^{-1}$ for $E_e$=300GeV option. The lower limit on $\beta_p^*$, which is given by proton bunch length, can be overcome by applying a "dynamic" focusing scheme [10], where the proton bunch waist travels with electron bunch during collision. In this scheme $\beta_p^*$ is limited, in principle, by electron bunch length, which is two orders magnitude smaller. More conservatively, an upgrade of the luminosity by a factor 3-4 may be possible. Therefore, luminosity values exceeding $10^{31}$cm$^{-2}$s$^{-1}$ seems to be achievable for al three options given in Table I.

Further increasing of luminosity can be achieved by increasing the number of protons in bunch and/or decreasing of normalised emittance. This requires the application of effective cooling methods at injector stages. Moreover, cooling in HERA ring may be necessary in order to compensate emittance growth due to intra-beam scattering [2]. First studies of the beam optics in the interaction region are presented in [1] and [2], where head-on collisions are assumed. The basic concept consists of common focusing elements for both the electron and the proton beams and separating the beams outside of the low-beta insertion. However, collisions with small crossing angle ($\approx$100 $\mu rad$) are also a matter of interest, especially for $\gamma p$ and $\gamma A$ options (see below). Further work on the subject, including the detector aspects, is very important.

## 3 MAIN PARAMETERS OF $\gamma p$ COLLIDER

Earlier, the idea of using high energy photon beams obtained by Compton backscattering of laser light off a beam of high energy electrons was considered for $\gamma e$ and $\gamma\gamma$ colliders (see [11] and references therein). Then the same method was proposed for constructing $\gamma p$ colliders on the base of linac-ring type $ep$ machines and rough estimations of the main parameters of $\gamma p$ collisions are given in [4]. The dependence of these parameters on the distance $z$ between conversion region and collision point was analyzed in [5], where some design problems were considered.

Referring for details to [5] let us note that $L_{\gamma p}=1.6\cdot 10^{31}$cm$^{-2}$s$^{-1}$ at $z$=10 $cm$ for TESLA$\otimes$HERA based $\gamma p$ collider with 300 GeV energy electrons beam. Here we take into account the electron-to-photon conversion coefficient 0.65 [11] and a factor of two coming from smallness of the photon bunch transverse sizes. Then, the luminosity slowly decreases with the increasing $z$ (factor ~1/2 at $z$=5 $m$) and opposite helicity values for laser and electron beams are advantageous (see Fig. 1). Additionally, a better monochromatization of high-energy photons seen by proton bunch can be achieved by increasing the distance $z$ (see Fig. 2). Finally, an upgrade of the luminosity by a factor 3-4 may be possible by applying a "dynamic" focusing scheme [10].

The scheme with non-zero crossing angle and electron beam deflection considered in [5] for $\gamma p$ option lead to problems due to intensive synchrotron radiation of bending electrons and necessity to avoid the passing of electron beam from the proton beam focusing quadrupoles. Alternatively, one can assume head-on-collisions (see above) and exclude deflection of electrons after conversion. In this case residual electron beam will collide with proton beam together with high-energy $\gamma$ beam, but because of larger cross-section of $\gamma p$ interaction the background resulting from $ep$ collisions may be neglected. The problem of over-focussing of the electron beam by the strong proton-low-$\beta$ quadrupoles is solved using the fact of smallness of the emittance of the TESLA electron beam. For this reason the divergence of the electron beam after conversion will be dominated by the kinematics of the Compton backscattering. In the case of 300 (800) GeV electron beam the maximum value of scattering angle is 4 (1.5) micro-radians. Therefore, the electron beam transverse size will be 100 (37.5) $\mu m$ at the distance of 25 $m$ from conversion region and the focusing quadrupoles for proton beam have negligible influence on the residual electrons. On the other hand, in the scheme with deflection there is no restriction on $n_e$ from $\Delta Q_p$, therefore, larger $n_e$ and bunch spacing may be preferable. All these topics need a further research.

Concerning the experimental aspects, very forward detector in $\gamma$-beam direction will be very useful for investigation of small $x_g$ region due to registration of charmed and beauty hadrons produced via $\gamma g \rightarrow Q^*Q$ sub-process [12].

## 4 MAIN PARAMETERS OF eA COLLIDER

The main limitation for this option comes from fast emittance growth due to intra beam scattering, which is approximately proportional $(Z^2/A)^2(\gamma_A)^{-3}$. In this case, the using of flat nucleus beams seems to be more advantageous because of few times increasing of luminosity lifetime. Nevertheless, sufficiently high luminosity can be achieved at least for light nuclei. For example, $L_{eC}=1.1\cdot 10^{29}$cm$^{-2}$s$^{-1}$ for collisions of 300 GeV energy electron beams and Carbon beam with $n_C=8\cdot 10^9$ and $\varepsilon_C^N=1.25\pi\cdot$mrad$\cdot$mm. This value corresponds to $L_{int}\cdot A \approx 10 pb^{-1}$ per working year ($10^7$ s) needed from the physics point of view [13]. Similar to the $ep$ option, the lower limit on $\beta_A^*$, which is given by nucleus bunch length, can be overcome by applying a "dynamic" focusing scheme [10] and an upgrade of the luminosity by a factor 3-4 may be possible.

## 5 MAIN PARAMETERS OF $\gamma A$ COLLIDER

In our opinion this is the most promising option of the TESLA$\otimes$HERA complex, because it will give unique opportunity to investigate small $x_g$ region in nuclear medium. Indeed, due to the advantage of the real $\gamma$ spectrum heavy quarks will be produced via $\gamma g$ fusion at characteristic

$$x_g \approx \frac{4m^2_{c(b)}}{0.9\times(Z/A)\times s_{ep}},$$

which is approximately $(2\div 3)\cdot 10^{-5}$ for charmed hadrons.

As in the previous option, sufficiently high luminosity can be achieved at least for light nuclei. Then, the scheme with deflection of electron beam after conversion is preferable because it will give opportunity to avoid limitations from $\Delta Q_A$, especially for heavy nuclei. The dependence of luminosity on the distance between conversion region and interaction point for TESLA$\otimes$HERA based $\gamma C$ collider is similar to that of the $\gamma p$ option [5]: $L_{\gamma C}=1.3\cdot 10^{29}$cm$^{-2}$s$^{-1}$ at $z$=0 and $L_{\gamma C}=10^{29}$cm$^{-2}$s$^{-1}$ at $z$=5 $m$ with 300 GeV energy electron beam. Let us remind that an upgrade of the luminosity by a factor 3-4 may be possible by applying a "dynamic" focusing scheme. Further increasing of luminosity will require the cooling of nucleus beam in the main ring. Finally, very forward detector in $\gamma$-beam direction will be very useful for investigation of small $x_g$ region due to registration of charmed and beauty hadrons.

## 6 MAIN PARAMETERS OF FEL $\gamma$A COLLIDER

Colliding of TESLA FEL beam with nucleus bunches from HERA may give a unique possibility to investigate "old" nuclear phenomena in rather unusual conditions. The main idea is very simple [7]: ultra-relativistic ions will see laser photons with energy $\omega_0$ as a beam of photons with energy $2\gamma_A\omega_0$, where $\gamma_A$ is the Lorentz factor of the ion beam. For HERA $\gamma_A = (Z/A)\gamma_p = 980(Z/A)$, therefore, the region $0.1 \div 10$ MeV, which is matter of interest for nuclear spectroscopy, corresponds to $0.1 \div 10$ keV lasers, which coincide with the energy region of TESLA FEL. The excited nucleus will turn to the ground state at a distance $l = \gamma_A \cdot \tau_A \cdot c$ from the collision point, where $\tau_A$ is the lifetime of the excited state in nucleus rest frame and $c$ is the speed of light. The huge number of expected events ($\sim 10^{10}$ per day for 4438 keV excitation of $^{12}$C) and small energy spread of colliding beams ($\leq 10^{-3}$ for both nucleus and FEL beams) will give opportunity to scan an interesting region with $\sim 1$ keV accuracy.

Table I. Upgraded TESLA Parameters

| Electron Energy, GeV | 300 | 500 | 800 |
|---|---|---|---|
| Number of e per bunch, $10^{10}$ | 1.8 | 1.5 | 1.41 |
| Number of bunches/pulse | 4640 | 4950 | 4430 |
| Beam power, MW | 40 | 30 | 40 |
| Bunch length, mm | 1 | 0.4 | 0.3 |
| Bunch spacing, ns | 192 | 192 | 192 |
| Repetition rate, Hz | 10 | 5 | 5 |

Table II. Upgraded Parameters of Proton Beam

| Beam energy, TeV | 0.92 |
|---|---|
| No. of particles/bunch, $10^{10}$ | 30 |
| No. of bunches/ring | 90 |
| Bunch separation, ns | 192 |
| $\varepsilon_p^N$, $10^{-6}$ $\pi\cdot$rad$\cdot$m | 0.8 |
| $\beta_p^*$, cm | 20 |
| $\sigma_{x,y}$ at IP, $\mu$m | 12.5 |
| $\sigma_z$, cm | 15 |

## CONCLUSION

It seems that HERA machine will not be the end point for lepton-hadron collisions. Linac-ring type ep and other colliders based on this will give an oppurtinity to go far in this direction. However more activity is needed.

## ACKNOWLEDGEMENTS

This work is supported by Turkish State Planning Organisation under the Grant No DPT-97K-120420 and DESY.

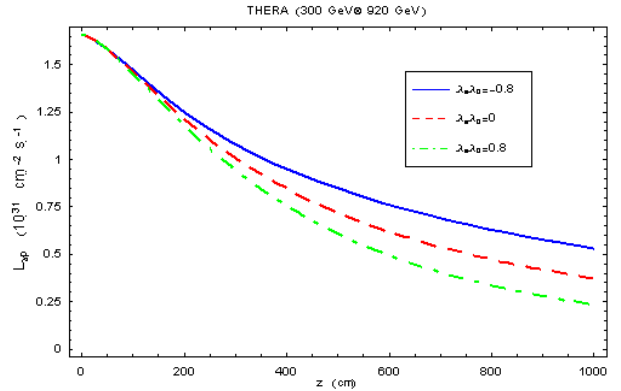

Figure 1.

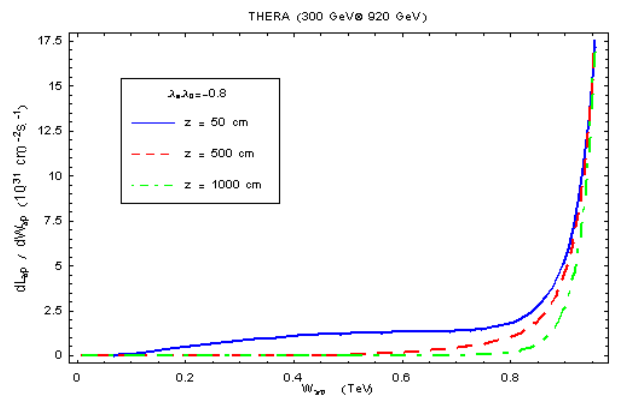

Figure 2.